\documentclass{aa}
\usepackage{graphicx}
\usepackage{natbib}
\usepackage{rotating}
\usepackage[varg]{txfonts}
\usepackage{color}
\usepackage{xspace}
\usepackage{upgreek}
\usepackage{hyperref}
\hypersetup{
    pdfborder={0 0 0 0},
    colorlinks=true,
    linkcolor=blue,
    urlcolor=blue,
    citecolor=blue}

\bibpunct{(}{)}{,}{a}{}{,}
\def\aap{A\&A}

\def\apjs{ApJS}

\def\apjl{ApJ}

\begin{document}

\title{High resolution radio imaging of the two Particle-Accelerating Colliding-Wind Binaries HD\,167971 and HD\,168112} 

\author{M.~De~Becker\inst{1} \and B.~Marcote\inst{2} \and T.~Furst\inst{1} \and P.~Benaglia\inst{3}}

\offprints{M.~De~Becker}

\institute{Space sciences, Technologies and Astrophysics Research unit -- STAR, University of Li\`ege, Quartier Agora, 19c, All\'ee du 6 Ao\^ut, B5c, B-4000 Sart Tilman, Belgium
\and
Joint Institute for VLBI ERIC, Oude Hoogeveensedijk 4, 7991 PD, Dwingeloo, The Netherlands
\and
Instituto Argentino de Radioastronomía (CCT La Plata, CONICET; CICPBA; UNLP), C.C.5, (1894) Villa Elisa, Buenos Aires, Argentina
}

\date{Received ; accepted }

\abstract
{The colliding-wind region in binary systems made of massive stars allows us to investigate various aspects of shock physics, including particle acceleration. Particle accelerators of this kind are tagged as Particle-Accelerating Colliding-Wind Binaries, and are mainly identified thanks to their synchrotron radio emission.}
{Our objective is first to validate the idea that obtaining snapshot high-resolution radio images of massive binaries constitutes a relevant approach to unambiguously identify particle accelerators. Second, we intend to exploit these images to characterize the synchrotron emission of two specific targets, HD\,167971 and HD\,168112, known as particle accelerators.} 
{We traced the radio emission from the two targets at 1.6 GHz with the European Very Long Baseline Interferometry Network, with an angular resolution of a few milli-arcseconds.}
{Our measurements allowed us to obtain images for both targets. For HD\,167971, our observation occurs close to apastron, at an orbital phase where the synchrotron emission is minimum. For HD\,168112, we resolved for the very first time the synchrotron emission region. The emission region appears slightly elongated, in agreement with expectation for a colliding-wind region. In both cases the measured emission is significantly stronger than the expected thermal emission from the stellar winds, lending strong support for a non-thermal nature.}
{Our study brings a significant contribution to the still poorly addressed question of high angular resolution radio imaging of colliding-wind binaries. We show that snapshot Very Long Baseline Interferometry measurements constitute an efficient approach to investigate these objects, with promising results in terms of identification of additional particle accelerators, on top of being promising as well to reveal long period binaries.}


\authorrunning{Authors}
\titlerunning{EVN imaging of HD167971 and HD168112}

\maketitle

\section{Introduction}\label{intro}

Massive stars, including O-type objects and their evolved Wolf-Rayet (WR) counterparts are known to produce strong stellar winds through the line-driving mechanism \citep{CAK,Puls2008}. These winds are made of a plasma at a temperature of a few 10$^4$\,K active at producing thermal Bremsstrahlung in the radio domain. The thermal radio spectrum of massive star winds is known to be optically thick, with a dependence of the flux density ($S_\nu$) as a function of frequency ($\nu$) such as $S_\nu \propto \nu^\alpha$, and a spectral index $\alpha$ close to 0.6 \citep{WB,PF}. However, some deviations with respect to this behaviour pointed to the existence of a complementary emission mechanism for a significant number of massive stars. This additional process is synchrotron radiation, and it is characterized by a spectral index $\alpha$ significantly lower than the thermal value, and even be definitely negative. Synchrotron radiation requires the existence of a magnetic field (of stellar origin), and a population of relativistic electrons. The latter is most likely produced by Diffusive Shock Acceleration (DSA, \citealt{Drury1983}). The occurrence of this process requires the presence of strong magneto-hydrodynamic shocks, and these are provided by the strong wind collisions that happen when two (or more) massive stars are part of a multiple system. Provided the stellar separation is high enough, the stellar winds collide at their terminal speeds, leading to velocity jumps of the order 2\,000--3\,000\,km\,s$^{-1}$. 
The subset of colliding-wind binaries able to accelerate particles up to relativistic velocities are referred to as Particle-Accelerating Colliding-Wind Binaries (PACWBs). To date, about 50 PACWBs have been identified, mainly thanks to the signature of synchrotron radio emission \citep{catapacwb}\footnote{An updated version of the catalogue is available at \url{http://www.astro.uliege.be/~debecker/pacwb/}.}.

The interest of studying PACWBs is mainly twofold. First, it allows to investigate shocks physics, including particle acceleration and non-thermal emission processes, in astrophysical laboratories different from supernova remnants (SNRs). Shock speed, local magnetic field and geometry are different, but physical processes are the same. Second, particle acceleration in pre-supernova massive star environment opens up the possibility of a moderate contribution to the production of cosmic rays. Even though SNRs are very likely the major contributors to galactic cosmic rays, one cannot reject the idea that a small contribution may come from alternative sources. The latter point calls upon an investigation of the population of particle accelerators, aimed at exploring the question of the fraction of PACWBs among massive binaries \citep{DeBecker2017,DeBeckerGCRBINA}. 

As PACWBs are composite emitters (thermal emission from individual winds and non-thermal emission from the colliding-wind region), it is not straightforward to disentangle their emission components. Most radio interferometers operated at centimetric wavelengths can reach angular resolutions of a few arcseconds, that is not enough to resolve the wind collision from the inner parts of the winds. The only way to spatially resolve the synchrotron emission from the thermal emission is the use of Very Long Baseline Interferometry, with baselines of thousands of kilometers. The angular resolution at centimetric wavelengths can be as good as a few milli-arcseconds (mas), therefore allowing to spatially resolve binary systems with orbital period of at least a few years at distances of a few kpc.

To date, only a handful of PACWBs benefited of such high angular resolution measurements. We can mention in particular WR\,147 \citep{williamswr147}, WR\,140 \citep{Doug140}, WR\,146 \citep{oconnorwr146art}, Cyg\,OB2\,\#9 \citep{vlbi2006}, Cyg\,OB2\,\#5 \citep{ortizcyg5}, HD\,93129A \citep{benaglia2015}, and more recently the most powerful synchrotron emitter in this category, Apep \citep{Marcote2021}.

This study aims at investigating the synchrotron emission region from two PACWBs, \object{HD~167971} and \object{HD~168112}. The idea is to make use of the European VLBI Network (EVN) to obtain valuable radio images of both systems, with the aim to characterize their synchrotron emission at the mas scale. Even though VLBI techniques have already been applied to HD\,167971 \citep{SanchezBermudez2019}, this paper is reporting on the very first high resolution radio imaging of HD\,168112. The paper is organized as follows. The two targets are introduced in Sect.\,\ref{Targets}. The data acquisition and processing are described in Sect.\,\ref{Data}. Our results are presented in Sect.\,\ref{sec:results} and discussed in Sect.\,\ref{Disc}. We finally summarize and conclude in Sect.\,\ref{Concl}.

\section{Targets: two well-established PACWBs}\label{Targets}

The two systems investigated in this paper are located in the NGC\,6604 open cluster \citep{reipurth6604}, at a distance of about 1.7\,kpc. Both are part of the catalogue of PACWBs \citep{catapacwb}, and have already been targets of several observation campaigns across the electromagnetic spectrum. 

HD\,167971 (BD-12 4980) is a hierarchical triple system made of a 3.32\,d period system (O6-7V + O6-7V) with a third object (O8I) on a wider close to 20-year orbit \citep{ibanoglu2013}. Adopting the naming convention proposed by \citet{debecker6604new}, the short period system is made of components Aa and Ab, and the third star is component B. After a first confirmation of an orbital motion between components A and B using infrared interferometry \citep{vlti167971}, the very first three-dimension characterization of the wide orbit has been achieved by a combination of interferometry and spectroscopy by \citet{LeBouquin2017}. The period of the wide orbit is 7\,803 $\pm$ 540\,d, with a significant eccentricity ($e$ = 0.44 $\pm$ 0.02).

The radio emission at centimetric wavelengths has been investigated in detail by \citet{Blo167971}, confirming the synchrotron emitter status through all main usual criteria (significant variability, non-thermal spectral index, high brightness temperature). HD\,167971 is the brightest synchrotron emitter among O-type PACWBs. Light curves present clear evidence for a strong variability of the radio emission compatible with the long period in the system. The maximum flux density was measured in 1988, that is coincident with the expected periastron passage according to the ephemeris published by \citet{LeBouquin2017}. This is in agreement with the expectation that the maximum of the synchrotron emission should occur close to periastron passage (see Sect.\,\ref{Disc167971}). More recently, \citet{SanchezBermudez2019} published the results of the first VLBI observations of HD\,167971 at two epochs, 2006 and 2016. 
\bigskip

HD\,168112 (BD–12 4988) displays strong hints for binarity as revealed by radio and X-ray measurements \citep{DeB168112}, even though no orbital solution exists at the time of writing this paper. Even the period of the system is still completely unknown. High angular resolution imaging in the visible could however identify the presence of a visual companion at an angular distance of about 3.3\,mas \citep{sana2014}. The flux ratio between the primary (O5.5III, component A) and the secondary allowed \citet{debecker6604new} to propose an O5.5--7.5III spectral type for the secondary (component B). On the basis of the variation of the X-ray emission produced in part by the wind-wind interaction region, a conservative lower limit on the eccentricity of 0.38 could also be determined \citep{debecker6604new}.

The radio emission from HD\,168112 shows clear evidence for synchrotron radiation, i.e. non-thermal spectral index and variability \citep{DeB168112,Blo168112}. However, this system has never been the target of high angular resolution imaging measurements before the present study.

\section{Observations and data reduction}\label{Data}

Our observations with the European VLBI Network (EVN) were performed on 5 November 2019 10:00--17:30~UTC at 18~cm (1.67 GHz) with project code EM131F (PI: B.~Marcote), with the following 15 participating stations: Lovell Telescope, Effelsberg, Westerbork single dish, Medicina, Noto, Onsala 32-m, Tianma, Urumqi, Toru\'n, Hartebeesthoek, Svetloe, Zelenchuskaya, Badary, Irbene, and Sardinia.
The data were correlated with the SFXC software correlator \citep{keimpema2015} at the Joint Institute for VLBI ERIC (JIVE; The Netherlands) with the total bandwidth of 256~MHz divided in eight 32-MHz subbands, with 64 spectral channels each, full circular polarization, and 2-s time integration.

The source J1743$-$0350 was used as fringe finder and bandpass calibrator. J1825$-$0737 was used as phase calibrator for both target sources, HD~167971 and HD~168112, in a phase-referencing cycle of 3.5~min on target and 1.5~min on the calibrator. The targets were alternated on each phase-referencing cycle. As a result, HD~167971 and HD~168112 were observed for a total of  $\approx 2~\mathrm{h}$ each. We note that the angular separation between the phase calibrator and the targets is $5.0^\circ$ and $4.8^\circ$, respectively.

The EVN data have been reduced in {\sc AIPS}\footnote{The Astronomical Image Processing System ({\sc AIPS}) is a software package produced and maintained by the National Radio Astronomy Observatory (NRAO).} \citep{greisen2003} and {\sc Difmap} \citep{shepherd1994} following standard procedures. \emph{A-priori} amplitude calibration was performed using the known gain curves and system temperature measurements recorded on each station during the observation as provided by the EVN Pipeline. We corrected for the ionosphere dispersive delays using the maps of total electron content provided by the Global Positioning System satellites via the {\tt TECOR} task in {\sc AIPS}.
We manually removed bad data (mainly frequencies and times affected by radio frequency interference). We first corrected for the instrumental delays and bandpass calibration using the fringe-finder source, and thereafter conduct the global fringe-fit of the data using all calibrator sources. The phase calibrator was then imaged and self-calibrated to improve the final calibration of the data in {\sc Difmap}. The obtained solutions were transferred to the target sources, which were finally imaged.

The cleaned images were obtained by model fitting a 2D Gaussian to the $(u,v)$ data. Due to the limited signal-to-noise ratio in the images and sparse $(u,v)$ coverage, this approach produced the optimal characterization of the sources while keeping a minimum number of free parameters. The resulting images did then allow us to characterize the geometry of the wind-collision region for each target.

\section{Results}\label{sec:results}

\begin{figure*}
    \includegraphics[width=\textwidth]{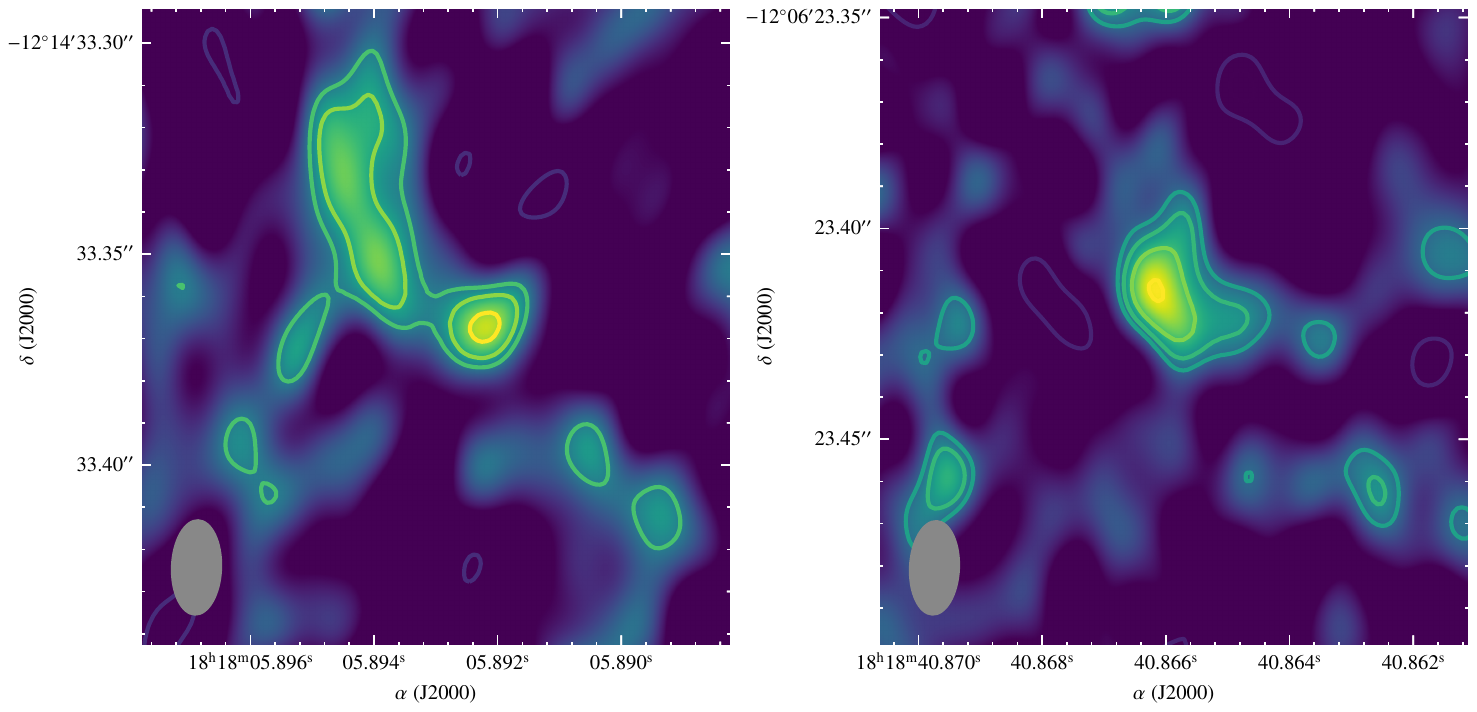}
    \caption{Field of HD~167971 (left) and HD~168112 (right) as seen by the EVN on 5 November 2019 at 18~cm. Both sources are detected in our data with a significance of $7\sigma$ and $9\sigma$ above the rms noise level of $140$ and $70\,\mathrm{\upmu Jy\ beam^{-1}}$, respectively. Contours start at $3\sigma$ the rms noise level and increase by factors of $\sqrt{2}$. The synthesized beam size of $12 \times 22~\mathrm{mas}^2,\ PA = 3^\circ$, is represented at the bottom left corner of each image as a gray ellipse.}
    \label{fig:evn-image}
\end{figure*}

Both sources, HD~167971 and HD~168112, are detected as compact sources at the mas scale in our observation on 5 November 2019. In the following, we present the results obtained from the aforementioned EVN data, independently for each target source. These results are also summarized in Table\,\ref{res} and the resulting images of both targets are shown in Figure~\ref{fig:evn-image}.

HD~167971 is detected as a compact radio source at the position (J2000) of $\alpha = 18^{\rm h}18^{\rm m}5.89300^{\rm s} \pm 1.1~\mathrm{mas},\ \delta = -12^\circ14^\prime33.3672^{\prime\prime} \pm 1.1~\mathrm{mas}$. We note that the quoted uncertainties represent the $1\sigma$ confidence interval and take into account the statistical uncertainties of the position measured in our data, the uncertainty in the position of the phase calibrator source, J1825$-$0737, within the international celestial reference frame (ICRF)\footnote{As obtained from the Radio Fundamental Catalog (RFC) version 2023C. See \url{http://astrogeo.org}.}, and the estimated uncertainties associated with the phase-referencing technique \citep{Kirsten2015}.

The radio source is detected with a peak brightness of $0.97 \pm 0.14~\mathrm{mJy\ beam^{-1}}$. A Gaussian fit to the emission reports a total flux density of $3.42 \pm 0.14~\mathrm{mJy}$ extended over a linear size of $34.8 \pm 0.3~\mathrm{mas}$ with a position angle of $78 \pm 7^\circ$.  The 2D Gaussian fit converted to a one dimensional Gaussian; i.e.\ the source size in one of the dimensions is negligible when compared to the resolution of the images ($12 \times 22~\mathrm{mas^2},\ PA = -2.7^\circ$). We note that the absolute amplitude calibration of the EVN data can exhibit uncertainties of $\sim 15\%$ due to the intrinsic calibration procedures related to VLBI arrays. These uncertainties were added in quadrature to the aforementioned value and the final flux density measurement is quoted in Table~\ref{res}.\\

HD~168112 is detected as a radio source with a flux density of $1,65 \pm 0.12~\mathrm{mJy}$ and peak brightness of $0.65 \pm 0.07~\mathrm{mJy\ beam^{-1}}$ (prior absolute uncertainties in the absolute flux density scale) at the position (J2000) of $\alpha = 18^{\rm h}18^{\rm m}40.86587^{\rm s} \pm 1.0~\mathrm{mas},\ \delta = -12^\circ6^\prime23.4233^{\prime\prime} \pm 1.0~\mathrm{mas}$.

A Gaussian fit to this source provided a measurement of the source size of $28.3 \pm 0.6~\mathrm{mas}$, elongated towards the direction $61.7 \pm 1.5^\circ$ (measured from the North direction to the East). The source size in the orthogonal direction was negligible when compared to the synthesized beam size.\\

Figure~\ref{fig:evn-image} shows the obtained images for both HD~167971 and HD~168112. In the case of HD~167971 we detect a compact radio source at the center of the map, while the elongated structure observed at its North-East part is ambiguously connected to the source. Indeed, the $3$--$4\sigma$ blobs detected in the South left and right corners of the images (similar in position for both images), are indicative of some phase losses and possible flux from the target that got smeared into the map. These structures actually disappear when performing a self-calibration of the data based on the fitted Gaussian model, indicative of not being astrophysical signals. This is not surprising in these data due to the following factors: the targets exhibit a low declination that made them visible only at low elevations for most of the EVN array. During the observation, both targets were observed at an average elevation of $\sim 20^\circ$. Furthermore, the phase calibrator is located at a significant separation with respect to the targets ($\sim 5^\circ$), making the phase-referencing technique less accurate when transferring the calibration from the phase calibrator to the targets (i.e.\ the assumption that all sources require the same corrections because the angular separation between them is negligible is slightly deviated). Finally, this observation was performed at a data rate of 2~Gbps (total bandwidth of 256~MHz) after the increase of bandwidth in the EVN. While the larger bandwidth is indeed beneficial to improve sensitivity, at the observed frequency (1.67~GHz), the presence of radio frequency interference (RFI) has significantly increased in the last years. At this band, several of the subbands were strongly affected by RFI, producing major changes in the gains of the telescopes\footnote{We note that the EVN no longer conduct observations at 256~MHz at 1.67~GHz because of this issue. At this band the total bandwidth is currently limited to 128~MHz to avoid the strong RFI.}. This resulted into a larger scatter in amplitudes than usual, producing the aforementioned artifacts in the images and a potential loss of part of the source fluxes. The final flux density values quoted in Table~\ref{res} take this into account. We clarify that these unexpected features, i.e. the North-East elongated structure along with the southern ones, do not appear at all in the radio images of HD\,167971 published by \citet{SanchezBermudez2019} on the basis of their VLBA data. Furthermore, these structures completely disappear when we perform a self-calibration on the data, confirming that they are originated by phase losses during the calibration process.
As a result of these artifacts in the data, the geometry in the source structures at the $3$--$4\sigma$ levels are not fully reliable. This is the reason why instead of conducting a fit assuming a bow-shaped structure as expected in these sources \citep[see e.g.][]{benaglia2015,Marcote2021}, we limited the analysis to a 2D Gaussian. This approach allows us to constrain the position angle of the emission, without being biased to possible curvature in the source structure that is not real.

\begin{table}
\caption{Summary of the properties of the radio sources (at 1.67 GHz) associated to the two observed targets on 5 November 2019. The coordinates are quoted for a J2000 epoch.\label{res}}
{\footnotesize
\begin{tabular}{l c c}
\hline
 & HD\,167971 & HD\,168112 \\
\hline

RA & $18^{\rm h}18^{\rm m}05.89300^{\rm s} \pm 1.1~\mathrm{mas}$ & $18^{\rm h}18^{\rm m}40.86587^{\rm s} \pm 1.0~\mathrm{mas}$ \\
Dec & $-12^\circ14^\prime33.3672^{\prime\prime} \pm 1.1~\mathrm{mas}$ & $-12^\circ06^\prime23.4233^{\prime\prime} \pm 1.0~\mathrm{mas}$ \\
$S_{{\rm 1.67~GHz}}$ & 3.4\,$\pm$\,0.7~mJy & 1.7\,$\pm$\,0.4~mJy \\
\hline
\end{tabular}}
\end{table}

\section{Discussion}\label{Disc}

\subsection{Synchrotron emission}\label{synch}
\subsubsection{HD\,167971}\label{Disc167971}

In this triple system, two colliding-wind regions are active: between components Aa and Ab, and between components A and B, respectively. Given the size of the orbit of the AaAb binary, it is very likely that any putative synchrotron emission from there would be completely suppressed by free-free absorption (FFA). However, the synchrotron emission that is measured certainly comes from the wide orbit, in agreement with the light curve obtained by \cite{Blo167971}. On the basis of the ephemeris published by \citet{LeBouquin2017} for the wide orbit, the orbital phase of our EVN measurement is $\phi = 0.52$. Our observation thus occurred very close to apastron. Not counting the effect of FFA, one expect the synchrotron emission to be at minimum at apastron. The reason is twofold. First, the local magnetic field in the emission region (of stellar origin) decreases as the distance between the stars increases \citep{UM}. Second, when the separation is longer the density of the colliding flows is lower, very likely leading to a drop in the injection rate of particles into the DSA process. One can thus expect our measurement to be at the minimum of the light curve. This is indeed confirmed when checking the light curve published by \citet{Blo167971}, that presents a minimum compatible with an expected time of apastron passage. As a result, the flux density from the synchrotron emission region at any other orbital phase should be greater than our 2019 measurement.

Our radio data alone are not enough to determine the position of the stars relative to that of the detected source. However, one can discuss some expectations based on the wind parameters of the two components. According to \citet{LeBouquin2017}, the projected semi-major axis of the 21-yr orbit is about 18.15\,mas. With an eccentricity of 0.443, this leads to a stellar separation at apastron of about 26\,mas, with a line of centers very close to the East-West direction. Based on the wind parameters quoted in Table\,\ref{adoptedparam}, the wind momentum rate ratio ($\eta = \dot{M_B}\,v_{\infty,B}\,\dot{M_A}^{-1}\,v_{\infty,A}^{-1}$) is about 0.38. This leads to an angular separation between the wind collision and component A (resp. B) of about 10\,mas (resp. 16\,mas). These numbers are a bit below and above the East-West width of the synthesized beam (12\,mas), respectively. In Fig.~\ref{fig:evn-image} (left), this translates into a potential position component A a bit beyond the left (East) limit of the second contour of the radio source, and component B should be a bit further on the right (West) of the source. However, we recommend some caution as these considerations are based on assumed wind parameters and not on a direct astrometric determination of the position of the stars.

A comparison with the measurements published by \citet{SanchezBermudez2019} is also relevant. One has to extrapolate flux densities at the same frequencies to proceed with a valid comparison. The 2006 and 2016 measurements are summarized in Table\,\ref{SB}. However, one has to note that the spectral indices at both epochs point to a surprisingly steep spectrum, much steeper than the expectation for a standard optically thin synchrotron spectrum from a population of relativistic electrons accelerated through DSA in high Mach number shocks, in the test-particle regime ($\alpha = -0.5$). On the basis of the spectral index map, \citet{SanchezBermudez2019} consider that an $\alpha$ value of about $-1.1$ is more typical of a significant part of the synchrotron emission region. They interpret the steepness of the synchrotron spectrum as a likely consequence of efficient inverse Compton cooling, leading to a softening of the relativistic electron spectrum. 

We also want to stress that such a steep spectral index could also be a signature of shock modification. Basically, the back-streaming of high energy particles upstream provides a contribution to the compression, leading to a drop in upstream velocity right behind the shock front. This creates a shock precursor that constitutes a main component of the modified shock structure. When relativistic electrons are back-scattered by scattering centers located in the precursor and not in the far-upstream region, they feel a lower velocity jump that translates into a lower compression ratio as compared to high Mach number adiabatic shocks for monoatomic gas  ($\chi$ < 4). The mean free path of relativistic particles diffusing in the magnetized plasma is proportional to the gyroradius, depending itself on particle energy. As a result, lower energy relativistic electrons diffuse across shorter distance upstream and do not go beyond the shock precursor. As the electron index ($p$) depends on the compression ratio, 
\begin{equation}\label{index}
p = \frac{\chi + 2}{\chi - 1},
\end{equation}
lower energy electrons are characterized by a greater $p$ value, leading to a relativistic electron population that is steeper than expected from linear DSA by unmodified shocks. In a given magnetic field, the typical synchrotron photon frequency is directly proportional to the relativistic electron energy. This back reaction of relativistic particles on the shock structure is thus expected to lead to a steeper synchrotron spectrum at low frequencies. This is typically what is observed in the case of young supernova remnants where particle acceleration efficiency is high enough to push DSA in the non-linear regime, leading to steep synchrotron radio spectra \citep{RadioSNR2015}.

\begin{table}
\caption{Summary of the measurements of the radio flux density of HD\,167971 published by \citet{SanchezBermudez2019}.\label{SB}}
\begin{center}
\begin{tabular}{crr}
& \multicolumn{1}{c}{2006} & \multicolumn{1}{c}{2016}\\
\hline\\[-8pt]
$S_{\rm 5.0~GHz}\ \mathrm{(mJy)}$ & $12.31 \pm 0.29$ & $7.32 \pm 0.18$\\
$S_{\rm 8.4~GHz}\ \mathrm{(mJy)}$ & $5.32 \pm 0.20$ & $2.82 \pm 0.16$\\[+1pt]
\hline\\[-8pt]
 $\alpha$\tablefootmark{a} & $-1.62 \pm 0.16$ & $-1.84 \pm 0.23$\\
\hline
\end{tabular}
\end{center}
\tablefoot{
\tablefoottext{a}{Error bar not published by \citet{SanchezBermudez2019}, calculated using Eq.\,1 in \citet{DeBecker2019}}
}
\end{table}

\begin{table*}
\caption{Adopted parameters and predicted thermal free-free flux densities for the two targets.\label{adoptedparam}}
\begin{center}
\begin{tabular}{l c c c c c c}
\hline
 & \multicolumn{3}{c}{HD\,167971} & & \multicolumn{2}{c}{HD\,168112}\\
 \cline{2-4}\cline{6-7}
Component & Aa & Ab & B & & A & B \\
Spectral type & O6--7V & O6--7V & O8I &  & O5.5III & O5.5--O7.5III \\
$\dot{M}~\mathrm{(M_\odot\,yr^{-1})}$ & $1.2 \times 10^{-7}$ & $1.2 \times 10^{-7}$ & $6.6 \times 10^{-7}$ &  & $1.13 \times 10^{-6}$ & $4.8 \times 10^{-7}$ \\
$v_\infty~\mathrm{(km\,s^{-1})}$ & 3\,000 & 3\,000 & 2\,800 &  & 2\,500 & 2\,400 \\ 
$T_{\rm eff}~\mathrm{(K)}$ & 36\,800 & 36\,800 & 31\,000 &  & 38\,000 & 35\,600 \\
$S_{\rm 1.67~GHz}~\mathrm{(mJy)}$ & $<0.001$ & $<0.001$ & $0.006$ &  & $0.015$ & $0.005$ \\
\hline
\end{tabular}
\end{center}
\tablefoot{
Stellar wind parameters are taken from \citet{predmdot}. Terminal velocities are estimated as 2.6 times the escape velocity. For components with uncertain spectral types, interpolated values have been adopted.
}
\end{table*}
Assuming a spectral index of $-1.1$, one can extrapolate the flux density we measured with the EVN to the bands used by \citet{SanchezBermudez2019}. We obtain flux densities of $\sim 1.0$ and $\sim 0.58~\mathrm{mJy}$ at 5.0 and 8.4 GHz, respectively. These values are indeed significantly lower than measured in 2006 and 2016 epochs, in agreement with the statement that our measurement occurs close to apastron, when the synchrotron emission should be minimum. One can consider that our measurement close to apastron is providing us with the bottom flux value of the full radio light curve.

Our radio measurement deserves to be compared to the expected thermal free-free emission from the winds of the system. We adopted the approach described for instance by \citet{WB} to calculate reasonable values of the thermal flux density for the three stellar winds contributing to this triple system. In agreement with \citet{leietal}, we assumed values of the molecular weight, the mean electron number, and the RMS ionic charge of 1.8, 1.0, and 1.0, respectively. Focusing on the outer parts of the wind where the measurable thermal radio emission is expected to come from, we assumed a clumping factor of the order of 4.0 \citep{Runacres2002} and an electron temperature equal to 30$\%$ of the stellar effective temperature \citep{Drew1990}. All useful parameters are summarized in Table\,\ref{adoptedparam}. We thus predict thermal flux densities at the $\mathrm{\upmu Jy}$ level for all component at 18~cm. These values are about three orders of magnitude lower than our actual measurement. Let's also mention that in principle some thermal emission is also expected to arise from the colliding-wind region \citep{PittardRadio2010}. The shocked wind material is a hot plasma likely to lead to some free-free emission. A significant thermal contribution from this region would require sufficiently high densities (emission process proportional to the square of the density) that could appear in very short binaries. However, for long period systems, the winds are colliding at a much longer distance, leading to a severe drop in the density that is scaling as $1/r^2$, with $r$ being the distance from the star to the wind collision. According to the simulations by \citet{PittardRadio2010}, one may expect a significant contribution from colliding wind regions of radiative shocks that occur in quite short period systems, especially at radio and sub-millimeter frequencies well above our measurement frequency. On top of that, one cannot call upon any free-free emission enhancement due to the clumpiness of the emitting material, as most likely density structures do not survive the shocks and dissipate in the post-shock region of adiabatic shocked winds \citep{Pittard2007}. At 1.67 GHz for a wide system producing adiabatic shocks such HD\,167971, the thermal emission arising from the colliding-wind region is certainly even lower than that coming from the individual winds. The radio source we measured with the EVN is thus by no means compatible with thermal emission. This is lending further support to the idea that our EVN source is attributable to the synchrotron emission region.

\subsubsection{HD\,168112}\label{Disc168112}

The lack of existing orbital solution for this system prevents us from interpreting our radio measurement as a function of the orbital phase. However, the striking and important result is the very first imaging of the synchrotron emission region. In the absence of any other VLBI measurement at other epochs and at other frequencies for this source, one has to be careful with the interpretation of the nature of the measured radio emission. In other words, one has to be sure that we are not dealing with thermal radio emission.

First of all, we predicted the expected thermal free-free emission flux density adopting the same approach as for HD\,167971 above. Our predictions lead to values of the order of $\sim 10~\mathrm{\upmu Jy}$ (see Table\,\ref{adoptedparam}). These values are much lower (almost a factor 200) than our measurement of $1.7 \pm 0.4~\mathrm{mJy}$. The predicted thermal emission is clearly below our detection threshold. As in the case of HD\,167971, we are dealing with a rather long period system producing adiabiatic shocks, leading therefore to a negligible putative thermal emission contribution from the stellar winds, especially at a frequency as low as 1.67\,GHz. Second, the morphology of the source is significantly elongated, with the elongation of the emission region not coincident with the major axis of the synthesized beam. This is in agreement with expectation that the synchrotron emission region should be somewhat extended, in coincidence with the colliding-wind region. Once again, this measured morphology is not compliant at all with the expectation of a purely unresolved, point source for thermal stellar winds. These two facts can fully be interpreted in terms of a synchrotron emission region at the limit to be resolved. These features are not at all compatible with thermal emission from the winds.

The measured position angle for this elongation ($61.7 \pm 1.5^\circ$) constraints the expected position for the two stars of  the system. We would thus expect these stars to be placed within a position angle (at the epoch of the observation, 5 November 2019) of $\approx 152^\circ$. Unfortunately, the angular separation between the stars cannot be constrained with these data.

\subsection{Energy budget}\label{enbud}

Given the negligible contribution of thermal emission at our measurement frequency, the radio flux densities reported here can be seen as clean and direct measurement of the synchrotron emission from the two targets. The synchrotron emission discussed in Sect.\,\ref{synch} actually results from a chain of energy conversion processes, on top of which lies the kinetic power of the stellar winds involved in the shock physics responsible for particle acceleration. The kinetic power, expressing the rate of transfer of mechanical energy by an individual stellar wind, can be expressed as
\begin{equation}\label{pkin}
P_\text{kin} = \frac{1}{2}\,{\dot M}\,v_\infty^2 = 3.155 \times 10^{35}\,{\dot M}_{-6}\,v_{\infty,8}^2\,\,\,\,\mathrm{(erg\,s^{-1})},
\end{equation}
\noindent where ${\dot M}_{-6}$ is the mass loss rate in units of 10$^{-6}$\,M$_\odot$\,yr$^{-1}$, and $v_{\infty,8}$ is the terminal velocity in units of 10$^8$\,cm\,s$^{-1}$. Using the wind parameters quoted in Table\,\ref{adoptedparam}, we determined the total kinetic power for both systems using Eq.\,\ref{pkin}, resulting from the sum of individual contributions from all components.  

\begin{table}
\caption{Energy budget.\label{enbudtable}}
\begin{center}
\begin{tabular}{l c c}
\hline
 & {HD\,167971} & {HD\,168112}\\
\hline
$P_\text{kin}$ (erg\,s$^{-1}$) & 2.3\,$\times$\,10$^{36}$ & 3.1\,$\times$\,10$^{36}$ \\
$L_\text{synch}$ (erg\,s$^{-1}$) & 3.0\,$\times$\,10$^{29}$ & 1.5\,$\times$\,10$^{29}$ \\
$RSE$ & 1.3\,$\times$\,10$^{-7}$ & 4.8\,$\times$\,10$^{-8}$ \\
$RSE_\text{lower}$ & 3.0\,$\times$\,10$^{-8}$ & 2.4\,$\times$\,10$^{-8}$ \\
$RSE_\text{upper}$ & 6.0\,$\times$\,10$^{-7}$ & 5.3\,$\times$\,10$^{-7}$ \\
\hline
\end{tabular}
\end{center}
\end{table}

The synchrotron luminosity results from the integration of the flux density in a given spectral domain, 
\begin{align}
L_\text{synch} & = 4\pi\,d^2\,\int_{\nu_\text{min}}^{\nu_\text{max}}\,S_\nu\,\text{d}\nu\nonumber\\
 & = 4\pi\,d^2\,S_{\nu,0}\,\frac{\nu_0^{-\alpha}}{\alpha + 1}\,\bigg[\nu_\text{max}^{\alpha + 1} - \nu_\text{min}^{\alpha + 1}\bigg]\,\,\,\,\mathrm{(erg\,s^{-1})},\label{lsynch}
\end{align}
\noindent where the flux density dependence on the frequency has been defined as $S_\nu \propto \nu^{\alpha}$, $d$ is the distance, and $\nu_0$ is the frequency of our measurement. The lower integration boundary can be set to 0.1\,GHz as it is very likely that the synchrotron emission will be self-absorbed below that value. For the upper boundary, a conservative value of 100\,GHz was adopted, as above that value the synchrotron flux is not expected to contribute much, given the negative index of the spectrum.

Assuming $\alpha = -0.5$ for a standard population of relativistic electrons accelerated by DSA in high Mach number shocks, we computed $L_\text{synch}$ using Eq.\,\ref{lsynch} for both systems and we obtained values of the order of a few times 10$^{29}$ erg\,s$^{-1}$ (see Table\,\ref{enbudtable}). One can then determine the radio synchrotron efficiency ($RSE$) defined as the fraction of the wind kinetic power converted into synchrotron radiation, i.e. $RSE = L_\text{synch}/P_\text{kin}$. Our results lie between the boundaries presented by \citet{DeBecker2017} and quoted as $RSE_\text{lower}$ and $RSE_\text{upper}$ in Table\,\ref{enbudtable}, in full agreement with expectations for the radio emission from PACWBs.

In the case of HD\,167971, as our measurement occurred close to apastron, the $L_\text{synch}$ and $RSE$ values quoted here should be considered as minimum values. In addition, in that part of the orbit the separation is so large that it is unlikely that the current measurement is affected by significant FFA. As a result, our energy budget estimate should be considered as representative of the actual synchrotron production at that orbital phase. For HD\,168112, the lack of any clear information on the orbit prevents us from interpreting further our results. Depending on the size and phase of the orbit (that is known to be quite eccentric, see Sect.\,\ref{Targets}), one may expect some free-free absorption to occur, especially close to periastron. We can thus not claim that our estimate of $L_\text{synch}$ (and $RSE$) is typical of the actual synchrotron radiation production. Our measured quantities should thus be considered as lower limits on the synchrotron radiation production rate at that specific epoch. 

For any specific PACWB at any orbital phase, $RSE$ depends on several energy conversion factors: (i) the fraction of the kinetic power actually invested in shock physics, (ii) the fraction of that energy injected into particle acceleration, (iii) the fraction of energy in non-thermal particles that is going to relativistic electrons, and finally (iv) the fraction of relativistic electron energy that is radiated as synchrotron radiation. Even though it sounds reasonable to consider that most of these energy conversion factors may be similar from one system to the other, they cover a parameter space that is large enough to allow $RSE$ to be significantly different among the population of PACWBs. A deeper investigation of these energetic considerations is required for a larger sample of systems, at various orbital phases and on the basis of high quality radio measurements to better quantify their potential to convert wind kinetic energy into synchrotron radiation.

\subsection{The synchrotron/binarity correlation}

Although this study is dedicated to the study of synchrotron emission from massive stellar systems, it is important to place it in a broader context. It is clear from the catalogue of PACWBs \citep[including its updated on-line version]{catapacwb}, that there is a strong correlation between the PACWB status (mainly based on the identification of synchrotron emission) and the binary (or higher multiplicity) status. On the one hand, this constitutes an incentive to search for new PACWBs focusing on already identified binaries. On the other hand, however, this opens up the possibility to make use of tracers of synchrotron radiation to establish the binary status of massive stars that are still elusive from the perspective of their multiplicity study using usual techniques. In other words, given that the so-called synchrotron/binarity coorrelation is quite well-established, synchrotron emission should also be considered as a tracer of multiplicity among massive stars.

In the framework, HD\,168112 is a quite instructive example. Despite some hints for binarity from its behaviour in X-rays \citep{debecker6604new} and the identification of an astrometric companion with no monitoring to trace a potential orbital motion to date \citep{sana2014}, this object has never reveal any radial velocity variation suggestive of an orbital motion. Adding the radio variation to this picture, we can talk about converging hints pointing to a binary status. The high angular resolution imaging we propose here, as a unique snapshot measurements, is actually enough to ascertain the binary status of the object.

This opens up the possibility to use the method presented in this paper to identify binaries among massive stars. However, one has to clarify that this approach is valid provided the synchrotron emission is not too much free-free absorbed by the stellar winds. This puts some significant constraints on the binary parameter space likely to be investigated. Basically, only sufficiently long period systems are relevant, i.e. with periods of at least several months, or even years in the case of thicker stellar winds of Wolf-Rayet stars responsible for FFA up to longer distances. This fact is also quite interesting, given that longest period systems are in general more difficult to identify (requirement for a long monitoring to identify variations, eccentric orbits showing significant radial velocity variations in a narrow range of orbital phase, slower orbital motion reducing the amplitude of the radial velocity curve...). Thus, provided that the high angular resolution radio measurement does not occur at a phase of strong FFA (i.e. periastron), it is likely to reveal features directly pointing to a binary status. Of course, a unique measurement is not enough to access a full description of the orbit, but this approach is worth being used to first identify yet unrevealed binaries, before organizing more focused campaigns using spectroscopic or astrometric techniques with the aim to characterize the orbit.

\section{Summary and conclusions}\label{Concl}

We reported on the results of VLBI measurements of two well-established PACWBs, members of the NGC\,6604 open cluster: HD\,167971 and HD\,168112. We used the European VLBI Network at 18 cm to image the radio emission from the two systems at one epoch, 5 November 2019. In both cases, we significantly detect a radio source, that we attribute to synchrotron radiation from the colliding-wind region. 

In the case of HD\,167971, our measurement is coincident with apastron, where the synchrotron radiation is at its lowest level. The source is unresolved, and too bright to be explained by thermal free-free emission from the stellar winds of the components of the system. For HD\,168112, we obtain the very first image of the synchrotron emission region. The radio source appears a bit elongated with a position angle significantly different from that of the synthesized beam, lending support to a slightly resolved synchrotron source. HD\,168112 becomes therefore the 8th PACWB whose synchrotron emission region has been imaged using high angular resolution radio measurements. The other systems are WR147, WR140, WR146, Cyg\,OB2\,\#9, Cyg\,OB2\,\#5, HD\,93129A, HD\,167971 and Apep.

Based on energy budget considerations, we quantified the synchrotron emission in terms of the fraction of the wind kinetic power effectively converted into synchrotron radiation. In the case of HD\,167971, the lack of significant FFA makes our estimate typical of the actual synchrotron radiation production rate. However, in the case of HD\,168112, the lack of information on the size and phase of the orbit prevents us from being more specific in our interpretation. In both cases, the mechanical-to-synchrotron energy conversion is fully compliant with expectations for PACWBs.

Finally, we stress the relevance and importance of snapshot high angular resolution radio imaging in the identification of synchrotron radio emitters. This is particularly true of HD\,168112, which displays signs of binarity, although it has not yet provided us with the ultimate proof. The clear detection of the radio source with a flux density much greater than expected from the thermal emission from stellar winds, along with a morphology deviating from that of an unresolved point source, is enough to confirm the synchrotron nature of the emission, that in turn is a clear sign a colliding-winds in a massive binary system. As a consequence, on top of being an efficient tool to identify synchrotron radio emission from massive stars, high angular resolution radio imaging constitutes a highly relevant method to unveil binary systems that are still elusive from the perspective of usual multiplicity study techniques. The latter point is especially relevant for the identification of long period binaries.

\begin{acknowledgements}
The authors would like to thank the referee for a fair report and constructive comments that helped to improve the paper. This research is part of the PANTERA-Stars collaboration, an initiative aimed at fostering research activities on the topic of particle acceleration associated to stellar sources\footnote{\url{https://www.astro.uliege.be/~debecker/pantera}}. The European VLBI Network is a joint facility of independent European, African, Asian, and North American radio astronomy institutes. Scientific results from data presented in this publication are derived from the following EVN project code: EM131. The SIMBAD database was used for the bibliography.
BM acknowledges financial support from the State Agency for Research of the Spanish Ministry of Science and Innovation under grant PID2019-105510GB-C31/AEI/10.13039/501100011033 and through the Unit of Excellence Mar\'ia de Maeztu 2020--2023 award to the Institute of Cosmos Sciences (CEX2019-000918-M).
\end{acknowledgements}

\bibliographystyle{aa}

\bibliography{bibmd}


\end{document}